\documentclass[fleqn,twoside]{article}
\usepackage{espcrc2}

\usepackage{graphicx}
\usepackage[figuresright]{rotating}
\usepackage{ifpdf}

\newcommand{\dncde}{\ensuremath{\mathrm{d}N_{\mathrm{ch}}/\mathrm{d}\eta}}
\newcommand{\pt}{\ensuremath{p_{\mathrm{t}}} }

\newcommand {\mass} {\mbox{\rm GeV$\kern-0.15em /\kern-0.12em c^2$}}
\newcommand {\tev} {\mbox{${\rm TeV}$}}
\newcommand {\gev} {\mbox{${\rm GeV}$}}

\newcommand {\mom} {\mbox{\rm GeV$\kern-0.15em /\kern-0.12em c$}}
\newcommand {\gmom} {\mbox{\rm GeV$\kern-0.15em /\kern-0.12em c$}}
\newcommand {\mmass} {\mbox{\rm MeV$\kern-0.15em /\kern-0.12em c^2$}}
\newcommand {\mmom} {\mbox{\rm MeV$\kern-0.15em /\kern-0.12em c$}}

\newcommand{\hz}{\,\mbox{${\rm Hz}$}}
\newcommand{\khz}{\,\mbox{${\rm kHz}$}}

\newcommand{\ms}{\mbox{${\rm ms}$}}

\newcommand{\mbyte}{\,\mbox{${\rm MB}$}}

\newcommand{\gbyteps}{\,\mbox{${\rm GB/s}$}}
\newcommand{\pp}{\mbox{pp}}
\newcommand{\PbPb}{\mbox{Pb--Pb}}

\newcommand{\Jpsi} {\mbox{J\kern-0.05em /\kern-0.05em$\psi$}\xspace}

\def\NIM#1#2#3{Nucl. Instr. Meth. {\bf #1}\ (#2)\ #3}
\def\IEEE#1#2#3{IEEE Trans. Nucl. Sci. {\bf #1}\ (#2)\ #3}

\def\subIEEE{submitted to IEEE Trans. Nucl. Sci.}

\usepackage[T1]{fontenc}
\usepackage[latin1]{inputenc}  

\newif\ifcomment
\commentfalse

\title{Real-time TPC Analysis with the ALICE High-Level Trigger}

\author{
  V. Lindenstruth\address[KIP]{Kirchhoff Institut für Physik, Im Neuenheimer Feld 227, D-69120 Heidelberg, Germany},
  C.~Loizides\addressmark[BE]\address[IKF]{Institut für Kernphysik Frankfurt, August-Euler-Str. 6, D-60486 Frankfurt am Main, Germany}\address[BE]{Department of Physics, University of Bergen, Allegaten 55, N-5007 Bergen, Norway}\thanks{Author's email: loizides@ikf.uni-frankfurt.de},
  D.~Röhrich\addressmark[BE],
  B.~Skaali\address[OS]{Department of Physics, University of Oslo, P.O.Box 1048 Blindern, N-0316 Oslo, Norway},
  T.~Steinbeck\addressmark[KIP],\\
  R.~Stock\addressmark[IKF],
  H.~Tilsner\addressmark[KIP],
  K.~Ullaland\addressmark[BE],
  A.~Vestb{\o}\addressmark[BE] and
  T.~Vik\addressmark[OS]\\
          for the ALICE Collaboration
}

\begin{document}

\begin{abstract}
% $Id: abstract.tex,v 1.2 2004/02/11 17:46:54 loizides Exp $

%The central detectors of the ALICE experiment at LHC will produce a
%foreseen data size of up to 75 MByte/event at an event rate <200 Hz
%resulting in a rate of ~15 GByte/sec. This exceeds the foreseen mass
%storage bandwidth of 1.25 GByte/sec by a factor of 15. Online processing
%of the data is necessary in order to select interesting (sub)events, or
%to compress data efficiently by modeling techniques.

%Processing the data requires a massive parallel computing system, the
%HLT system located in the data flow after the front-end electronics of
%the detectors. The system will consist of a farm of clustered SMP-nodes
%based on off-the-shelf PCs connected with a high bandwidth low latency
%network. The system nodes will be interfaced to the front-end
%electronics via optical fibers connecting to their internal PCI-bus,
%using a custom PCI Receiver Card. These boards provide a FPGA
%co-processor for data intensive repetetive task of the pattern
%recognition.

%Most of the local pattern recognition will be done using the FPGA
%co-processor while the data is being transferred to the memory of the
%corresponding nodes. Algorithms for conventional cluster finding and
%local track finding based on a Circle Hough Transformation of the raw
%data are currently under development. Tests on prototypes are being
%done, using both the foreseen software for online data analysis and
%communication.  Latest results concerning the hardware and software
%implementation will be shown.

The ALICE High-Level Trigger processes data online, to either
select interesting (sub-) events, or to compress data efficiently by
modeling techniques. %Data intensive repetitive local pattern
%recognition tasks will be done in custom hardware.
Focusing on the main data source, the Time Projection Chamber,
the architecure of the system and the current state of 
the tracking and compression methods are outlined.

%two pattern recognition methods under investigation: a
%sequential approach (\emph{cluster finder} and \emph{track follower})
%and an iterative approach (\emph{track candidate finder} and
%\emph{cluster deconvoluter}). We show, that the former is suited for
%pp and low multiplicity PbPb collisions, whereas the latter might be
%applicable for high multiplicity PbPb collisions of dN/dy$\mathbf{>}$3000.
%if it turns out that more than charged particles would have 
%to be reconstructed inside the TPC. 
%Based on the developed tracking schemes we show that using
%modeling techniques a compression factor of around 10 might be
%achievable. 
%Benchmarks of the former indicate, that we can reach the designed
%inspection rate for pp collisions of 1 kHz with more than 98\% and
%for low multiplicity PbPb collisions of 200 Hz with more than 90\%
%track detection efficiency. Though, the method will clearly fail for
%high multiplicity central PbPb collisions, if more than $\sim$8000
%charged particles are produced in the TPC. Here the latter, iterative
%approach will be applicable with an total efficiency of more than
%90\%.  

\vspace{1pc}
\end{abstract}

% typeset front matter (including abstract)
\maketitle

% $Id: introduction.tex,v 1.4 2004/02/12 18:15:01 loizides Exp $

\section{Introduction}
\label{introduction}

The ALICE experiment described in~\cite{alice,aliceadd1,aliceadd2}
will investigate \PbPb\ collisions at a center of mass energy of 
about 5.5\,\tev\ per nucleon pair and \pp\ collisions at 14\,\tev. 
Its detectors are optimized for heavy-ion reactions 
at an anticipated charged particle multiplicity of up 
to $\dncde$ of 8000 in the central region~\cite{aliceppr1}.

The main central tracking detector, the Time Projection Chamber
(TPC), is read out by about 600\,000 channels, producing at most 
a data size of 75\,\mbyte\ per event for central 
\PbPb\ under extreme assumption and of 
2.5\,\mbyte\ for \pp\ collisions. 
%The drift time amounts to 88\,\musec; 
The estimated value of the maximum 
gate frequency is about 200-1000\,\hz\ depending 
on event multiplicity~\cite{alicetpc}.

The overall event rate is limited 
by the Data Acquisition (DAQ) 
bandwidth to the permanent storage system 
of 1.25\,\gbyteps. Without further reduction or 
compression the ALICE TPC can only take
central \PbPb\ events at up to 20\,\hz.
%\ and min.\,bias\footnote{A minimum bias trigger selects events with as
% little as possible bias in respect to the nuclear cross section.} \pp\
%events at up to 500\,Hz. 
Significantly higher rates are possible by either selecting 
interesting (sub-) events, or compressing data efficiently 
by modeling techniques; both requiring real-time analysis of
the detector information with a latency of the order 
of a few \ms. To accomplish the pattern recognition 
tasks at an incoming date rate of 10-25\gbyteps, a 
massive parallel computing system, 
the High-Level Trigger (HLT) system,
is being designed~\cite{hltdaqtrig}. 

%\subsection{Functionality}
%\label{functionality}
%The HLT system is intended to reduce the data rate produced by the
%detectors as far as possible to have reasonable taping costs. 

The key requirement for the HLT system is the ability to perform
the event analysis in real-time. Based on the extracted information, 
charge clusters and tracks, data reduction can be performed in different ways: 

\begin{itemize}
\item {\bf Trigger}: Generation and application of a software trigger
  capable of selecting interesting events from the input data stream. 
\item {\bf Select}: Reduction in the size of the event data by
  selecting (sub-) events and/or region of interest (RoI). 
\item {\bf Compression}: Reduction in the size of the event data by
compression techniques.  
\end{itemize}

As such the HLT system will enable the ALICE TPC 
detector to operate at a rate up to a few hundred \hz\ 
for heavy-ion and up to 1\,\khz\ 
for \pp\ collisions. %todo ref to physics? %\cite{hltphysics}

%In order to increment the statistical significance of rare
%processes, dedicated triggers can select candidate events or
%sub-events. By analyzing tracking information from the different
%detectors and (pre-)triggers  online, selective or partial readout of
%the relevant detectors can be performed thus reducing the event rate.

%The tasks of such a trigger are selections based upon the online
%reconstructed track parameters of the particles, e.g. to select events
%containing e$^+$e$^-$ candidates coming from quarkonium decay or to
%select events containing high energy jets made out of collimated beams
%of high $p_T$ particles~\cite{hltphysics}. In the case of low
%multiplicity events such as for pp collisions, the online
%reconstruction can be used to remove pile-up (superimposed) events
%from the trigger event. 

% $Id: architecture.tex,v 1.4 2004/02/12 18:15:01 loizides Exp $

\section{Data Flow and HLT Architecture}
\label{architecture}

\begin{figure}[htb!p!f]
\begin{center}
\includegraphics[width=7cm]{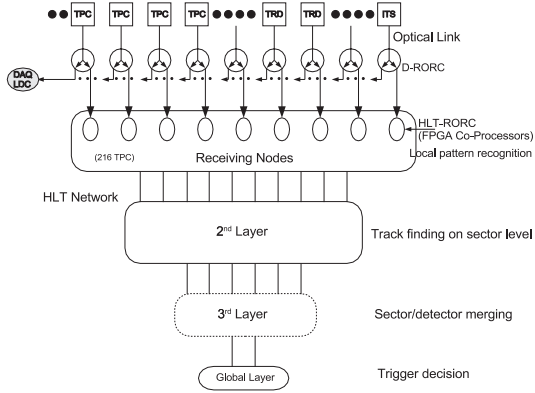}
\end{center}
\caption{Data-flow architecture of the HLT system. The detector 
raw data is duplicated and received by the DAQ and HLT system.}
\label{hltarch}
\end{figure}

Figure~\ref{hltarch} shows a sketch of the architecture of the
system adapted to the anticipated data flow from the ALICE
detectors, focusing on the TPC. 
The TPC consists of 36 sectors, 
each sector being divided into 6 sub-sectors. 
For each sub-sector the data  is transferred 
via an optical fiber from the detector front-end 
into DAQ Read-Out Receiver Cards (D-RORC). 
From there a copy of the data is sent 
into HLT Read-Out Receiver Cards (HLT-RORC) interfaced 
to the receiving nodes through the internal PCI bus. 
In addition to different communication interfaces, 
the HLT-RORC provides a FPGA co-processor for the data intensive 
local tasks of the pattern recognition and enough external 
memory to store several dozen event fractions. 
The receiver nodes perform the local
pre-processing task, cluster and track
seeding on the sub-sector level using 
the FPGA co-processor. The next two levels 
of computing nodes exploit the local neighborhood: 
track segment reconstruction on sector level. Finally all 
local results are collected from the other sectors 
(or in principle from the other ALICE sub-detectors)
and combined on a global level: track segment
merging and final track fitting. 

The computing farm is designed 
to be completely fault tolerant 
avoiding single points of failure, except for the unique 
detector links. A generic communication framework has 
been developed based on the publisher-subscriber 
principle, which one allows to construct any hierarchy 
of communication processing 
elements~\cite{pubsub,thesistimm}.

% $Id: patternrecognition.tex,v 1.4 2004/02/11 17:46:54 loizides Exp $

\section{Online Pattern Recognition}
\label{onlinepatternrecognition}

%Concerning the TPC, which is the main data source of ALICE delivering
%up to 15 GB/sec, the HLT system has to perform online pattern
%recognition at an event rate of $\le$ 200 Hz for PbPb collisions and
%$\le$ 1 kHz for pp collisions. Up to $\sim$ 16000 charged particle
%tracks per PbPb event have to be reconstructed within a time budget
%of 5 ms and up to $\sim$ 1000 for pp in 1 ms.  
In general, the HLT system has to reconstruct the complete event 
information online. Concerning the TPC and the other tracking 
devices, the particles ideally follow helical trajectories due to the
solenoidal magnetic field of the L3 magnet surrounding these central
detectors.
%Thus, we mathematically describe a track by a helix with 
%5(+1) parameters\footnote{To describe an arbitrary helix in 3 dimensions, 
%one needs 7 continuous parameters and a handedness switch. For the 
%special case of the ALICE geometry there are then 5 independent parameters 
%plus the handedness switch.}.
%For track fitting only the 5 parameters are relevant as the handedness 
%will be deduced by the particle's path.
%A TPC track is composed out of clusters. The pattern recognition task 
%for the HLT system is to process the raw data in order to find clusters 
%and to assign them to tracks thereby determining the helix track parameters 
%using different fitting strategies.

For HLT tracking, we distinguish two different approaches: the
``sequential feature extraction'' and the ``iterative feature
extraction''~\cite{tns03,thesisasv}.  

The sequential method approximates the cluster 
centroids using a fast ``Cluster Finder''. The centroids
are used as input for the ``Track Follower'' to determine 
the corresponding track parameters. This approach is applicable for lower 
occupancy like \pp\ and low multiplicity \PbPb\ collisions. 
However, at larger multiplicities expected for \PbPb\ at
LHC, clusters start to overlap and deconvolution 
becomes necessary in order to achieve the desired 
tracking efficiencies. 

Whereas, the iterative approach first estimates potential tracks
using a ``Track Candidate Finder'', which are then feeded to a 
``Cluster Fitter'' to assign clusters to the tracks
thereby deconvoluting overlapping clusters shared by 
different tracks candidates. In both cases, a final helix fit on 
the assigned clusters determines the track parameters. 

In order to reduce data shipping and communicaton overhead within the
HLT, as much as possible of the \emph{local} pattern recognition will
be done using the FPGA co-processor~\cite{gautechep03}. 

%--------------------------------------------------------------

\subsection{Sequential Tracking Approach}
\label{seqtracker}

\begin{figure}[thb!f]
\centering
\includegraphics[width=7cm]{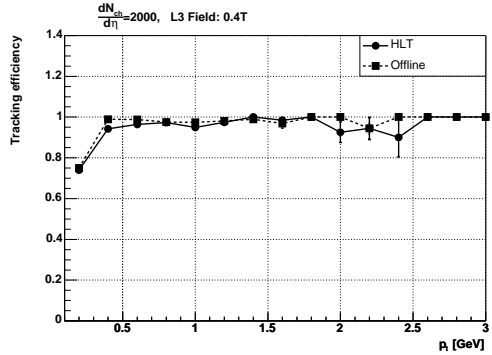}
\caption{Tracking efficiency for $\dncde=2000$
and a magnetic field strength of 0.4\,T as a function of $\pt$
compared for online (HLT)  sequential and offline tracking scheme.}
\label{eff04}
\end{figure}

The HLT tracking scheme for low multiplicity events has been 
adapted from the STAR L3 trigger~\cite{startrigger}.
The cluster finder estimates cluster centroids 
by the weighted mean in pad-and-time direction. Overlapping 
clusters are split at local minima. The list of space points 
is then handed to the track follower, which at first 
forms neighboring clusters into segments 
and then merges segments into tracks.
The tracking performance has been extensively studied 
and compared with the offline TPC reconstruction 
chain~\cite{hltdaqtrig,tns03,thesisasv}. It turns out, that 
the method is well suited and within the anticipated
time budget for charged particle multiplicities 
of up to $\dncde=2000$.

\begin{figure}[tbh]
\centering
\includegraphics[width=7cm]{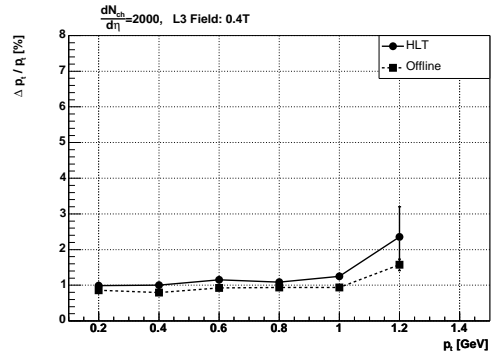}
\caption{Relative transverse momentum resolution for $\dncde=2000$
and a magnetic field strength of 0.4\,T as a function of $\pt$
compared for online (HLT)  sequential and offline tracking scheme.}
\label{ptres04}
\end{figure}

Concentrating on $\dncde=2000$ and a magnetic field of
0.4\,T, Fig.~\ref{eff04} shows a comparison of the tracking 
efficiency as a function of $\pt$ for the HLT and offline 
tracking schemes. In both cases, the efficiency drops 
for low momentum tracks ($\pt < 0.5\,\gev$) due to the 
crossing of sector boundaries. The integral efficiency is about 
90\% and the fake track rate of the order of 1.5\%. 
Fig.~\ref{ptres04} displays the momentum resolution as 
a function of $\pt$ for the same setup. On average, the 
HLT cluster finder takes about 15\,\ms, the tracker 
about 750\,\ms.

\subsection{Iterative Tracking Approach}
\label{ittracking}

The simple sequential method may be applicable to 
$\dncde=4000$, but ALICE is preparing  for even 
higher multiplicities of up to $\dncde=8000$. 
In that case, the occupancy in the TPC is of the order of 25\% 
and deconvolution becomes mandatory in order to achieve the
desired tracking efficiencies. The cluster shape is dependent
on the track parameters, and in particular on the crossing
angles of tracks between the padrow and drift time. 
In order to properly deconvolute the overlapping clusters, 
knowledge of the track
parameters that have produced the clusters is needed. 
Once track candidates are given, the clusters can be fit
to their known shape, and the cluster centroid can 
be correctly reconstructed. If the fit fails, the track
candidate will be disregarded.
%The cluster deconvolution is geometrically local, and thus trivially
%parallel, and could be performed in parallel on the raw data. 

%--------------------------------------------------------------

\subsubsection*{Hough Transform}

%The Hough transform is a standard tool in image analysis that allows
%recognition of global patterns in an image space by recognition of
%local patterns (ideally a point) in a transformed parameter space. The
%basic idea is to find curves that can be parameterized in a suitable
%parameter space. 

The Hough Transform (HT) method could be suitable 
to estimate track candidates, as it can be applied directly on 
the raw ADC data. In its original form~\cite{hough59} the HT
determines a curve in parameter space for a signal 
corresponding to all possible tracks with a given 
parametric form to which it could possibly belong. 
The space is then discretized and entries are stored in a
histogram. For all entries exceeding a given threshold 
in the histogram the corresponding parameters are found.

%As mentioned above, in ALICE the local track model is a helix. 
For simplification of the transformation the detector is divided 
into subvolumes in pseudo-rapidity. In addition, restricting
the analysis to tracks originating from the vertex, 
a track in the $\eta$-volume is characterized by two parameters: 
the emission angle with the beam axis $\psi$ and the curvature 
$\kappa$. The transformation is performed from (R,$\phi$)-space 
to ($\psi$,$\kappa$)-space using the following equations: 
\begin{eqnarray*}
R &=&\sqrt{x^2+y^2} \\
\phi &=& \arctan(\frac{y}{x}) \\
\kappa &=& \frac{2}{R}\sin(\phi - \psi)\,. \\
\end{eqnarray*}

%Each ADC value above a certain threshold transforms into a sinusoidal
%line extending over the whole $\psi$-range of the parameter space. All
%the corresponding bins in the histogram are incremented with the
%corresponding ADC value. The superposition of these point
%transformations produces a maximum at the circle parameters of the
%track. The track recognition is now done by searching for local maxima
%in the parameter space. 

%\begin{figure}[hbt]
%\centering
%\includegraphics[height=8cm,width=4cm,angle=-90]{pics/hough_eff_8000.eps}
%\caption{Tracking efficiency for the Hough transform 
%on a high occupancy event.%The overall efficiency is above 90\%.}
%\label{hougheff8000}
%\end{figure}

The integral efficiency for ``good'' track candidates is above 
90\% for a full multiplicity event and a magnetic field 
of 0.2\,T. This was estimated~\cite{tns03} by dividing the
number of verified track candidates divided with the number of
findable tracks estimated by the offline analysis . 
The list of verified track candidates was obtained by laying out 
a road in the raw data corresponding to the track parameters of
the peak. If enough clusters were found along the road, the track
candidate was considered a track, if not the track candidate was
disregarded. 
%However, one of the problems encountered with the Hough transform
%algorithm is the number of fake tracks coming from spurious 
%peaks in the parameter space. 
However, depending on certain threshold parameters, the
number of fake track candidates coming from spurious 
peaks in the parameter space is above 200\%. 
That problem of the HT method has to 
be overcome in order to determine 
valuable track candidates as input for the 
cluster fitting and deconvoluting procedure.

\subsubsection*{Cluster Fitter}

%todo anders write some more sentences?
The cluster fitting method were initially implemented for TPC 
data in the NA49 experiment. 
The Cluster Fitter (CF) fits a two-dimensional Gauss 
function with 5 parameters to the charge clusters:
\begin{itemize}
\item the position in pad and time direction;
\item the widths in pad and time;
\item the amplitude of the distribution.
\end{itemize}

The initial values of the fit parameters are provided by the track
parameters. The position in pad and time direction is obtained by
calculating the crossing point between the tracks and the
two-dimensional padrow plane. The widths are obtained from the
parameterisation of the cluster model as a function of the track
parameters. During the minimisation procedure of the least square
error, the widths are held fixed at their input values, while 
the position in pad and time are free to vary. 
%The fitting routine is an implementation of the Levenberg-Marquardt
%method~\cite{hltbib:marq}, which is a standard algorithm for the
%minimisation of the least square error in the fit of nonlinear models.
%Figure~\ref{HLTfig:hough_fit_example} shows an example of three
%overlapping clusters being unfolded by the fitting procedure.

\begin{figure}[htbp!f]
\centering
\ifpdf
\includegraphics[width=7cm]{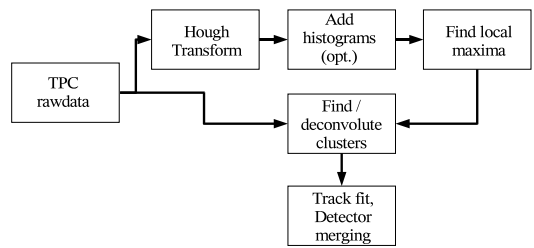}
\else
\includegraphics[bb=70 110 702 405, width=7cm]{./pics/houghflowdia}
\fi
\caption{Flow diagram showing the pattern recognition scheme 
         using the Hough Transform and the Cluster Deconvoluter/Fitter.} 
\label{houghflowdia}
\end{figure}

A flow diagram of the complete chain of HT and
CF is shown in Fig.~\ref{houghflowdia}. The
HT is done locally on the receiver nodes. The data volume therefore
corresponds to 1/6 of a complete TPC sector.
Optionally the respective histograms within a complete sector 
can be added in order to improve the signal/noise ratio of the 
peaks in the parameter space. A simple over-threshold local-maxima 
finder processes the histograms and identifies the
peaks corresponding to track candidates. The list of track candidates
is passed to the CF to deconvolute the clusters
along the respective track roads. The obtained tracks and their
assigned clusters are finally fitted to obtain their track parameters.

%\begin{figure}[htbp]
%\centerline{\includegraphics[width=8cm]{fit_example}}
%\caption{Example of deconvolution of overlapping clusters.
%  The deconvolution is done by fitting the charge distribution to
%  (in this case) three 2D gauss functions. The initial values of the
%  fit parameters are obtained by the track parameters, and the widths of
%  the distributions are held fixed at their input values.}
%\label{HLTfig:hough_fit_example}
%\end{figure}

\begin{figure}[htbp]
\centering
\includegraphics[width=7cm]{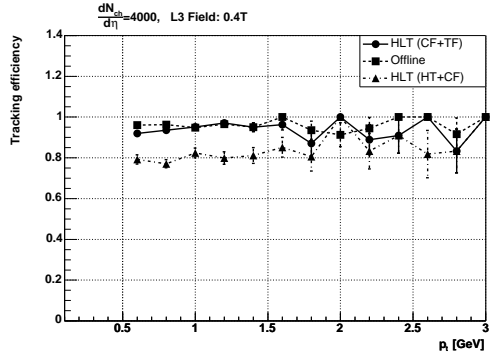}
\caption{Tracking efficiency of 
 the Hough Transform and Cluster Fitter (HT+CF) 
 as a function of $\pt>0.5\,\gev$ for $\dncde=4000$
 and a magnetic field strength of 0.4\,T.
 For comparison, the efficiency of 
 offline and the HLT sequential approach (CF+TF) 
 is also shown.}
 \label{highpteff}
\end{figure}

Figure~\ref{highpteff} shows the efficiency obtained using the 
HT and cluster fitting approach for $\dncde=4000$ and a magnetic
field strength of 0.4\,T as a function of $\pt$. %The tracking efficiency
%was defined in the same way as described in the previous section. 
For comparison, the results for the sequential cluster finder/track 
follower approach and the offline tracking scheme are added. 
The efficiency loss with respect to the standard tracking is 
significant. The reason is, that even for $\pt>0.5\,\gev$ 
the HT produces a large number of fake track candidates 
from structured backgrounds in the parameter 
space (as mentioned above). %Before the cluster fitting 
%is performed, a large  number of track candidates turn out to be fake. 
These fake tracks are filtered out during the cluster fitting
procedure by removing tracks, which do not point to valid clusters.
However, their presence causes the cluster fitting procedure to become
unstable, if too many fake tracks are pointing to the same cluster.
As a result, the cluster fit may be rejected on the basis of
$\chi^2$ criteria, resulting in a failure to fit the cluster. 

In order to ensure the principal function of the CF, 
offline tracks were feeded in the cluster fitting procedure 
representing optimal track candidates. Here, even at $\dncde=8000$ 
and a magnetic field strength of 0.4\,T, the cluster fitter 
was able to improve the relative $\pt$ resolution by 10\%
and the azimuthal resolution by 15\%, showing that the 
deconvolution/fitting procedure is working and even
capable of improving the offline resolution.

% $Id: compression.tex,v 1.5 2004/02/12 18:15:01 loizides Exp $

\section{Data modeling and Data compression}

Besides triggering and selecting (parts of) events, 
one of the mains tasks of the HLT system is to compress 
the event data efficiently with a minimal loss of physics information.
In general, two schemes of data compression are considered: 
\begin{itemize}
\item {\bf Binary lossless data compression}, allowing bit-by-bit
reconstruction of the original data set;
\item {\bf Binary lossy data compression}, not allowing bit-by-bit
reconstruction of the original data, while however retaining all
relevant physical information.
\end{itemize}

%Methods such as Run-length encoding (RLE), Huffman and LZW are 
%considered lossless compression, while thresholding and hit finding 
%operations are considered lossy techniques that could lead to a loss of small
%clusters or tail of clusters. It should be noted that data compression
%techniques in this context should be considered lossless 
%from a physics point of view. 
Most state of the art compression techniques were studied on
real NA49 and simulated TPC data and presented in detail 
in~\cite{berger02}. All methods roughly result 
in compression factors close to 2.

However, the most effective 
data compression can be expected by cluster and track 
modeling~\cite{tns03,thesisasv}. Here, the input to the compression 
algorithm is a lists of tracks and their
corresponding clusters. For every assigned cluster, the
cluster centroid deviation from the track model is calculated in both
pad and time direction. Its size is quantized with respect to the
given detector resolution and represented by a fixed number of bits. 
The quantization steps have been set to 0.5\,mm for the pad direction and 
0.8\,mm for the time direction, which is compatible with 
the intrinsic detector resolution.
In addition the total charge of the cluster is stored.
Since the cluster shape itself can be parameterised as a
function of track parameters and detector specific parameters, the
cluster widths in pad and time are not stored for every
cluster. During the decompression step, the cluster centroids are
restored, and the cluster shape is calculated based on the track
parameters. %In tables~\ref{trackparams} and~\ref{clusterparams}, 
%the track and cluster parameters are listed together with 
%their respective size being used in the compression. Instead of
%assigning only found clusters and their padrow numbers to a track,
%we store for every padrow a cluster structure with a minimum size
%of one bit, indicating whether the cluster is ``present'' or not.

%\begin{figure}[thb]
%\centering
%%\includegraphics[width=7cm]{pics/compress_eff.eps}
%\caption{Comparison of the tracking efficiency of the offline
%reconstruction chain before and after data compression. A total loss
%of efficiency of $\sim$1\% was observed.}
%\label{compresseff}
%\end{figure}

%\begin{figure}[thb]
%\centering
%%\includegraphics[width=7cm]{pics/compress_res.eps}
%\caption{Comparison of the $p_T$ resolution of the offline reconstruction
%chain before and after data compression.}
%\label{compressres}
%\end{figure}

The compression scheme has been applied to a simulated \PbPb\ 
event with a multiplicity of $\dncde=1000$. The input tracks used for the
compression are tracks reconstructed with the sequential 
tracking approach. The clusters, which were not assigned 
to a track during the track finding step,
were disregarded and not stored for further analysis.
%\footnote{The remaining clusters mainly originate from 
%very low $p_T$ tracks such as $\delta$-electrons, which 
%could not be reconstructed by the track
%finder. Their uncompressed raw data amounts to 
%a relative size of about 20\%.}. 
A relative size of 11\% for the compressed data with respect 
to the original set is obtained, whereas having a 2\% efficiency
loss compared to the original data.

%In order to evaluate the impact on the physics observables, 
%the compressed data is decompressed and the restored cluster 
%are processed by the offline reconstruction chain. 
%In Fig.~\ref{compresseff} the offline tracking efficiency before 
%and after applying the compression is compared as a function 
%of $p_T$. A total loss of about 2\% in efficiency is observed.
%Fig.\,\ref{compressres} shows for the same events the $p_T$ resolution 
%as a function of $p_T$ before and after the compression is applied.
%The observed improvement of the $p_T$ resolution is connected
%to way the errors of the cluster are calculated. For the case of
%the standard offline reconstruction chain the errors are calculated
%using the cluster information itself, whereas for the compression scheme
%they are calculated using the track parameters. 

%Keeping the potential gain of statistics by the increased 
%event rate written to tape in mind, one has to weight 
%the tradeoff between the impact on the physics observables 
%and the cost for the data storage. For occupancy events of
%more than 20\% (corresponding to $dN/dy>2000$), 
%clusters start to overlap and has to be properly deconvoluted 
%in order to effectively compress the data. 

%In this scenario, the Hough transform or another effective iterative
%tracking procedure would serve as an input for the cluster
%fitting/deconvolution algorithm. With a high online tracking
%performance, track and cluster modeling, together with noise removal,
%can reduce the data size by a factor of 10. 

% $Id: conclusions.tex,v 1.5 2004/02/12 18:15:01 loizides Exp $

\section{Conclusion}

Focusing on the TPC, the HLT system is designed to increase 
the readout and storage of relevant physics events by a 
factor of 10. 
The sequential approach, 
cluster finding followed by track finding, 
is applicable for pp and low multiplicity \PbPb\ data up to 
$\dncde=2000$ with more than 90\% efficiency and might
be extensible for higher particle densities.
The timing results show that the desired rate of 
1\,\khz\ for \pp\ and 200\,\hz\ for \PbPb\ will be 
achievable. 
For multiplicities of $\dncde \ge 4000$ ,
the iterative approach is foreseen using a track candidate 
finder followed by cluster deconvolution using the cluster fitter.
Here, the Circle Hough Transform as a possible method
to find track candidates seems to fail, as its high number
of fake track candidates confuses the otherwise well working
cluster fitter. So far, the problem to find a quantitative 
much better track candidate finder has not been solved. 
%Different solutions are investigated. 
%Among these are a combined approach where standard
%tracking performed in the outer padrows, where the occupancy is low,
%is combined with a HT in the inner padrows. In this way, 
%the track segments from the outer parts of the detector might be
%used as a filter for the Hough parameter space, which serves to reduce
%the complexity of the peak finding procedure and consequently minimise
%the number of false positives.

By compressing the data using data modeling techniques, the results 
for low multiplicity events show that with a loss of 
tracking efficiency of about 2\% a compression factor of
up to 10\% with respect to the original data sizes is achievable.

\end{document}